# Security Header Fields in HTTP Clients

Pascal Gadient, Oscar Nierstrasz
Software Composition Group, University of Bern
Bern, Switzerland
🌐 scg.unibe.ch/staff

Mohammad Ghafari
School of Computer Science, University of Auckland
Auckland, New Zealand
✉ m.ghafari@auckland.ac.nz

*Abstract*—HTTP headers are commonly used to establish web communications, and some of them are relevant for security. However, we have only little information about the usage and support of security-relevant headers in mobile applications. We explored the adoption of such headers in mobile app communication by querying 9 714 distinct URLs that were used in 3 376 apps and collected each server's response information. We discovered that support for secure HTTP header fields is absent in all major HTTP clients, and it is barely provided with any server response. Based on these results, we discuss opportunities for improvement particularly to reduce the likelihood of data leaks and arbitrary code execution. We advocate more comprehensive use of existing HTTP headers and timely development of relevant web browser security features in HTTP client libraries.

## I. INTRODUCTION

The most prominent clients for the HTTP protocol are web browsers. Modern web browsers receive regular updates every few weeks[1,2] and protect users from threats with various techniques such as HTTP headers, *e.g.*, HTTPS enforcement (HSTS) or continuous HTTPS certificate revocation feeds (certificate transparency). However, the support for such security header fields in HTTP client libraries, *e.g.*, `HTTPUrlConnection` in *OpenJDK* is missing.

We study the presence of HTTP header fields in the communication of mobile apps to understand their use and the provided protection. Using the dataset from Gadient *et al.* [1], we accessed 9 714 distinct URLs from 3 073 closed-source and 303 open-source apps. We issued for each URL an HTTP GET request to the server in order to trigger an HTTP(S) response. Each collected server response consists of the HTTP header (*e.g.*, storing the connection properties) and the HTTP body (which holds the user data, *e.g.*, plain text). We investigated the client-side support of the found header fields to understand the research question, *What is the support of the most common security-related HTTP header fields in existing HTTP clients?* In particular, we present the security-related HTTP header fields, their purpose, and prevalence, and we investigate which are supported in common HTTP libraries.

In the top 50 used header fields in the communication of mobile apps, we could identify sixteen well-known security-related fields. We found that on average 93% of the security-enabling pairs are not used in server responses. We discovered that all commonly used HTTP clients in Android apps lack proper support for the majority of such header fields. We discuss these header fields and explain how HTTP client libraries can benefit from them too. We also propose two new HTTP header fields to facilitate security.

The support for these headers requires changes in the configurations of server side software (*e.g.*, against version leaks), or in both the server and client side code (*e.g.*, for the data persistence policy), which remains to be studied in future work. We publicly share our dataset to encourage further research in this direction.[3]

In the remainder of this paper, we provide the necessary background in section II before presenting the used methodology in section III. We investigate the current HTTP header support in section IV and present our proposals in section V. We explain the threats to validity in section VI, and discuss related work in section VII. We conclude this paper in section VIII.

## II. BACKGROUND

We introduce the terms and technologies that are relevant to this work.

### A. Hypertext Transport Protocol (HTTP)

HTTP has initially been planned as a general purpose "application-level protocol for distributed, collaborative, hypermedia information systems" [2], but it is best-known for its delivery of content to web browsers. HTTP and its successor HTTP/2 provide facilities to encapsulate user data, *e.g.*, HTML, JSON, XML, or SOAP, and use plain-text messages to instruct the receiver on how to treat the transmitted data. HTTP Secure (HTTPS) is an extension of HTTP and thus follows the same principles, except that the messages are encrypted. As shown in Listing 1, requests and responses mostly follow the same structure, but there exist minor differences: the request always specifies the HTTP method (line 1), *e.g.*, `GET`, `POST`, `PUT`, `DELETE`, and the requested fully qualified resource path, *i.e.*, lines 1 and 2. On the contrary, the server response includes an HTTP status code (line 11) to indicate whether the request was successful, but not any resource path.

---

[1]Firefox Release Calendar,
https://wiki.mozilla.org/Release_Management/Calendar
[2]Google Chrome Roadmap,
https://www.chromestatus.com/features/schedule
[3]https://figshare.com/s/c57bb34cadcac225cadc

## B. HTTP header fields

In HTTP communication, header fields are used to set up the connection between the server and the client, *e.g.*, by specifying the used data encoding (line 6) and content caching option (line 9), and to provide additional information, *e.g.*, the used infrastructure (lines 3 and 14) or the originating date of the response (line 15). Besides non-standard fields, there exist for the sake of interoperability 48 different header fields in the HTTP/1.1 specification; each header field consists of a key-value pair in textual form. For example, the header field key `Content-Type` (line 12) declares the content type of the message body, *e.g.*, the value `text/plain` is used for plain text, `text/html` for websites, or `application/json` for JSON web API responses [2].

```
1  GET /v2/networks/nextbike-leipzig HTTP/1.1
2  Host: api.citybik.es
3  User-Agent: Mozilla/5.0 (Windows NT 10.0)
4  Accept: text/html,application/xhtml+xml
5  Accept-Language: en-US,en;q=0.5
6  Accept-Encoding: gzip, deflate
7  Connection: keep-alive
8  Upgrade-Insecure-Requests: 1
9  Cache-Control: max-age=0
10
11 HTTP/1.1 200 OK
12 Content-Type: application/json
13 Content-Length: 5613
14 Server: nginx/1.15.9 (Ubuntu)
15 Date: Mon, 14 Oct 2019 09:44:16 GMT
16 Access-Control-Allow-Origin: *
17 X-RateLimit-Limit-minute: 180
18 X-RateLimit-Remaining-minute: 179
19 X-Kong-Upstream-Latency: 61
20 X-Kong-Proxy-Latency: 1
21 Via: kong/1.2.1
```

Listing 1. Typical request and response communication flow between a client and a server

## C. Security-related Header Fields

Header fields can pose a threat, *e.g.*, leak software version information, or they can mitigate a threat, *e.g.*, code execution, click-jacking, and data leak.

Version information leaks are typically caused by `Server`, `X-Powered-By`, `X-AspNet-Version`, and `X-Powered-By-Plesk` header fields. These headers are prevalent in web communication and the risk is that when adversaries know about the software version, they can research that software *e.g.*, find publicly announced vulnerabilities in old software versions, and accordingly plan attacks against the web servers.

Header fields such as `X-Content-Type-Options`, `X-XSS-Protection`, or `Content-Security-Policy` can mitigate code execution attacks. Code execution attacks require two steps: the arbitrary injection of malicious code into an app, and its execution to steal sensitive data or to manipulate a rendered website. `X-Frame-Options` mitigates potential click-jacking attacks that can be performed *e.g.*, when several iframes are shown simultaneously, but one creates a view that overlays the others. This attack confuses users to accidentally click on clickable elements such as buttons and hyperlinks. The other fields we investigate in this work contribute to data leak prevention, which can mitigate unauthorized access to sensitive resources like a credential or insecure encryption.

## III. METHODOLOGY

For the analysis we used an URL list[4] to connect to servers and retrieve their responses, which we finally exercised. Consequently, the preliminary sourcing of the apps and the extraction of the URLs has been performed by the authors of the corresponding tool. Based on our results, we manually investigated the HTTP client support for security-related header fields.

### A. Sourced Apps

The URLs in the dataset are extracted from random Android apps found in the Google Play Store (3 073 apps) and the F-Droid repository (303 apps) that request Android's `INTERNET` permission. Based on their package identifier, only the most recent version of each app has been kept in the dataset. The Play Store apps come from 48 different categories, the most prevalent categories being `EDUCATION` (317 apps) and `TOOLS` (292 apps). Moreover, the apps have an average star rating of 4.2 stars (median: 4.3 stars), and most apps achieved between 100 and 1 000 downloads. Barely any app was downloaded more than one million times, and most of the apps were updated in 2018.

### B. URL Extraction

The static analysis tool used to extract the URLs performed three steps for each analyzed app. First, it decompiled the source code that is distributed within the APK installation file to regular Java code. Next, it detected the used web communication APIs in the code and extracted the corresponding data, *e.g.*, the tool can assemble URLs from concatenated string variables within a class, and even reconstruct JSON data structures from JSON object class implementations. Finally, the key-value pairs in the identified URLs were enriched with possible values, which could be found in the code, or their value type if no value was available. This process was error-prone and time demanding. A 32-core machine with 128 GB RAM was allowed to work on each app for up to 30 minutes before the process was killed, yet the analysis did not complete for some of them. After the analysis, the reported URLs were collected and duplicated URLs removed. URLs that point to the same server, but use a different path or query parameter were considered different to not miss any particular server configuration. In the end, 1 230 open-source URLs and 8 486 closed-source URLs were available for further analyses.

### C. Header Data Collection

The reported URLs consequently represent different kinds of HTTP servers, *e.g.*, for web APIs, media streaming, or website delivery. For every reported URL we issued an HTTP `GET` request and collected the response header information.

---
[4]available online, see footnote 3

An empty file has been created for those URLs for which we received no response in the process. After we collected the header information, we used simple pattern matching to gather the prevalence of the different response header fields.

*D. HTTP Client Support*

In order to understand whether an HTTP client supports the security-related header fields that we discovered, we searched each field name in the source code, the project website, documentation, and the forums where available. If we encountered matches, we started a manual investigation to determine the extent to which the support is available. For the web browser compatibility, we searched Mozilla's developer network directory, which provides browser compatibility matrices, and for non-standardized fields we had to use Google to find the relevant information. This task was performed by one of the authors and required about fifteen hours.

IV. RESULTS

In this section we first discuss the prevalence of header fields in server responses for mobile apps, before we show our findings regarding their support in non-browser HTTP clients.

*A. Identified Header Fields*

We found 439 header fields, which we could identify in the server responses. We present the top 50 in Table I. The first column denotes the rank among the most prevalent header fields, the second column denotes the number of occurrences in our dataset, the third column presents the header field name, and the fourth column reveals the purpose. We collected this information from Mozilla's Developer Network,[5] or, if unavailable, from websites operated by the corresponding protocol designers that could be found using Google Search. We proceeded identically for the last column, which shows the relevance to security for each header field together with a brief explanation. We highlighted the header fields that are relevant to security.

We found that the most prevalent header fields serve eight different purposes, *i.e.*, 16 (30%) allow *performance optimization*, 14 (26%) support *debugging*, 12 (23%) address *security*, 4 (8%) perform *advertisement*, 4 (8%) request *data presentation*, one (2%) enables *cookie management*, one allows *content redirection*, and, finally, one clarifies *privacy*. As we can see, only a subset of them are security-related, *i.e.*, 16 header fields (30%). Moreover, only four of them (8%) do not increase security, but pose a threat by leaking information.

*B. Security-related Header Fields*

Within the top 50 fields, we identified sixteen (30%) that may introduce or resolve a security threat, *e.g.*, fields that can prevent arbitrary code execution, click-jacking attacks, or data leaks. We considered fields that leak version information as also being security-related, because such data can make existing servers an easy target for adversaries. We present the details in Table II. The first and second columns denote the header field name and its intended use. The third column cites the relevant specification document, where available. The fourth column indicates either the corresponding threat that the field mitigates, or the threat that a field can introduce, *i.e.*, a version leak. The fifth column shows the total number of URLs returning a specific header field in the order `total responses (open-source responses / closed-source responses)`, and the last column reveals affected percentage of URLs following the same order.

From a total of 9 714 responses, the `Server` header field is omnipresent and included in 72% of them. We observe that `X-Powered-By`, `X-AspNet-Version`, and `X-Powered-By-Plesk` are less frequently adopted fields, which were present in 18.0%, 6% and 3% of the responses, respectively. The fields `X-Content-Type-Options` and `X-XSS-Protection` (both 16%), and `X-Frame-Options` (13%) are among the top five most used header fields, but they only occurred in around every seventh response. `Strict-Transport-Security` and `Expect-CT` are rarely used (8% and 3%). Nevertheless, whether these fields are relevant for every app is not known and needs to be investigated in future research.

> **Observation**
>
> The use of header fields that leak software information is common and such fields are enabled by default. The adoption of fields that improve security is rather rare. Nevertheless, not every field is relevant within an app context and further investigation is needed.

When comparing open and closed-source app URLs, we observed that open-source apps are slightly better. For example, the `Strict-Transport-Security` field was more common in open-source app responses, even by considering the more extensive use of HTTPS connections. Future work may investigate the underlying reasons for such a difference. Similarly, some well-known security-related HTTP header fields were more prevalent in open-source server responses, *e.g.*, `X-Content-Type-Options`, `Content-Security-Policy`, `X-Frame-Options`, and `X-XSS-Protection`.

*C. HTTP Client Support*

We present our findings in Table III. For a feature (*i.e.*, a security-related header field) that is fully supported we use the symbol ✔, and for those with limited support (✔). Limited support refers to features that only have partial support, *e.g.*, only selected options are implemented, the implementation has not yet been released, or the corresponding logic is only available as stub, *i.e.*, the most frequent header fields[6] are parsed, but not evaluated. We use the symbol ✘ for unsupported features. The following software releases have

---

[5]https://developer.mozilla.org/en-US/docs/Web/HTTP/Headers

[6]https://tools.ietf.org/html/rfc7541#appendix-A

TABLE I
TOP 50 HTTP HEADERS IN MOBILE APP WEB COMMUNICATION

| Rank | # Occ. | Header field | Purpose | Relevant to security |
|---|---|---|---|---|
| 01 | 7 567 | Date | performance optimization | Minor: provides a timestamp |
| 02 | 7 189 | Content-Type | data presentation | Minor: is a CORS-safelisted response header |
| *03* | *6 978* | *Server* | *advertisement* | *Major: can leak sensitive information* |
| 04 | 4 032 | Content-Length | performance optimization | Minor: is a CORS-safelisted response header |
| 05 | 3 479 | Cache-Control | performance optimization | Minor: is a CORS-safelisted response header |
| 06 | 3 065 | Connection | performance optimization | Minor: provides connection-state |
| 07 | 2 400 | Expires | performance optimization | Minor: is a CORS-safelisted response header |
| 08 | 2 111 | Set-Cookie | cookie management | Minor: cookie transmission |
| 09 | 1 811 | Vary | performance optimization | Minor: enables fine-grained caching |
| 10 | 1 788 | Location | content redirection | Minor: redirect target |
| *11* | *1 770* | *X-Powered-By* | *advertisement* | *Major: can leak sensitive information* |
| *12* | *1 601* | *X-Content-Type-Options* | *security* | *Major: can prevent content sniffing from arbitrary data* |
| *13* | *1 519* | *X-XSS-Protection* | *security* | *Major: can prevent XSS attacks* |
| 14 | 1 367 | Accept-Ranges | performance optimization | Minor: enables partial downloads |
| *15* | *1 289* | *X-Frame-Options* | *security* | *Major: can prevent iframe attacks* |
| 16 | 1 241 | Pragma | performance optimization | Minor: is a CORS-safelisted response header |
| 17 | 971 | Last-Modified | performance optimization | Minor: is a CORS-safelisted response header |
| *18* | *916* | *Access-Control-Allow-Origin* | *security* | *Major: extends cross origin resource sharing* |
| 19 | 834 | Etag | performance optimization | Minor: document identifier |
| *20* | *787* | *Strict-Transport-Security* | *security* | *Major: can prevent HTTPS downgrade attacks* |
| 21 | 659 | Alt-Svc | performance optimization | Minor: HTTP/2 load balancing |
| *22* | *601* | *Upgrade* | *security* | *Major: upgrades connection protocol or security* |
| 23 | 594 | P3P | privacy | Minor: privacy web page |
| *24* | *538* | *X-AspNet-Version* | *advertisement* | *Major: can leak sensitive information* |
| *25* | *517* | *Content-Security-Policy* | *security* | *Major: can restrict access to particular origins* |
| 26 | 471 | Via | debugging | Minor: routing information |
| 27 | 425 | X-Cache | debugging | Minor: caching state at the CDN |
| 28 | 410 | CF-Ray | debugging | Minor: request identifier |
| *29* | *336* | *Access-Control-Expose-Headers* | *security* | *Major: exposes selected headers to a frontend* |
| *30* | *332* | *Expect-CT* | *security* | *Major: enforces certificate transparency (obsolete)* |
| 31 | 318 | Access-Control-Allow-Methods | performance optimization | Minor: lists supported HTTP methods |
| 32 | 300 | X-UA-Compatible | data presentation | Minor: lists compatible user agents |
| 33 | 258 | Age | performance optimization | Minor: proxy cache duration |
| *34* | *255* | *X-Powered-by-Plesk* | *advertisement* | *Major: can leak sensitive information* |
| 35 | 250 | Access-Control-Allow-Headers | performance optimization | Minor: lists supported HTTP headers |
| 36 | 225 | Status | debugging | Minor: server response status |
| *37* | *205* | *Access-Control-Allow-Credentials* | *security* | *Major: exposes credentials to a frontend* |
| 38 | 196 | Transfer-Encoding | performance optimization | Minor: specifies data encoding |
| 39 | 192 | X-GitHub-Request-Id | debugging | Minor: request identifier |
| *40* | *189* | *Timing-Allow-Origin* | *security* | *Major: can introduce side-channel attacks on personalized data* |
| *41* | *186* | *Referrer-Policy* | *security* | *Major: can restrict the exposed referrer information* |
| 42 | 185 | X-Amz-CF-Id | debugging | Minor: request identifier |
| | 185 | X-Amz-CF-Pop | debugging | Minor: server identifier |
| 43 | 182 | X-Request-Id | debugging | Minor: request identifier |
| 44 | 175 | X-Cache-Hits | debugging | Minor: server side caching statistics |
| 45 | 174 | X-Served-By | debugging | Minor: lists CDN caching servers |
| 46 | 166 | X-FB-Debug | debugging | Minor: request debugging information |
| 47 | 163 | Allow | performance optimization | Minor: lists supported HTTP methods |
| 48 | 131 | Content-Disposition | data presentation | Minor: specifies data presentation |
| | 131 | X-Amz-Id-2 | debugging | Minor: request identifier |
| | 131 | X-Amz-Request-Id | debugging | Minor: request identifier |
| 49 | 127 | Content-Language | data presentation | Minor: is a CORS-safelisted response header |
| 50 | 123 | X-Timer | debugging | Minor: message transport statistics |

TABLE II
SECURITY-RELATED HTTP HEADER FIELDS FOUND IN SERVER RESPONSES SORTED BY THEIR PREVALENCE

| Header | Use | Specification | Threat | # Responses | % URLs |
|---|---|---|---|---|---|
| Server | Software used by the origin server to handle the request. | RFC 2616 [2] | version leak | 6 978 (909 / 6 069) | 70% (74% / 72%) |
| X-Powered-By | Specifies the technology supporting the web application. | non-standard | version leak | 1 770 (95 / 1 675) | 18% (8% / 20%) |
| X-Content-Type-Options | Can be used to require checking of a response's "Content-Type" header against the destination of a request. | WHATWG [3] | code execution | 1 601 (330 / 1 271) | 16% (27% / 15%) |
| X-XSS-Protection | Stops pages from loading when they detect reflected cross-site scripting (XSS) attacks. | non-standard | code execution | 1 519 (321 / 1 198) | 15% (26% / 14%) |
| X-Frame-Options | Indicates a policy that specifies whether the browser should render the transmitted resource within a <frame> or an <iframe>. | RFC 7034 [4] | click-jacking | 1 289 (317 / 972) | 13% (26% / 11%) |
| Access-Control-Allow-Origin | Indicates whether a resource can be shared based by returning the value of the Origin request header, "*," or "null" in the response. | WHATWG [3] | data leak | 916 (141 / 775) | 9% (11% / 9%) |
| Strict-Transport-Security | Indicates to a UA that it MUST enforce the HSTS Policy in regards to the host emitting the response message containing this header field. | RFC 6797 [5] | data leak | 787 (251 / 536) | 8% (20% / 6%) |
| Upgrade | Intended to provide a simple mechanism for transitioning from HTTP/1.1 to some other protocol on the same connection. | RFC 7230 [6] | data leak | 601 (0 / 601) | 6% (0% / 7%) |
| X-AspNet-Version | A state server implementation indicates which version of the state server is using this response header. | non-standard | version leak | 538 (38 / 500) | 5% (3% / 6%) |
| Content-Security-Policy | Preferred mechanism for delivering a policy from a server to a client. | W3C [7] | code execution | 517 (176 / 341) | 5% (14% / 4%) |
| Access-Control-Expose-Headers | Indicates which headers can be exposed as part of the response by listing their names. | WHATWG [3] | data leak | 336 (23 / 313) | 3% (2% / 4%) |
| Expect-CT | Allows web host operators to discover misconfigurations in their certificate transparency deployments. | IETF [8] | data leak | 332 (147 / 185) | 3% (12% / 2%) |
| X-Powered-By-Plesk | Advertises the used Plesk server software. | non-standard | version leak | 255 (0 / 255) | 3% (0% / 3%) |
| Access-Control-Allow-Credentials | Indicates whether the response can be shared when request's credentials mode is "include." | WHATWG [3] | data leak | 205 (7 / 198) | 2% (1% / 2%) |
| Timing-Allow-Origin | Defines an interface for web applications to access the complete timing information for resources in a document | W3C [9] | data leak | 189 (16 / 173) | 2% (1% / 2%) |
| Referrer-Policy | While the header can be suppressed for links with the noreferrer link type, authors might wish to control the Referer header more directly. | W3C [10] | data leak | 186 (79 / 107) | 2% (6% / 1%) |

been evaluated in this study: *Glide 4.7.0* from FEB-2018, *HttpComponents 5.1* from AUG-2021, *Ion* from NOV-2020, *LoopJ 1.4.9* from JAN-2021, *OkHttp* from AUG-2021, *RetroFit* from AUG-2021, *Volley 1.2.1* from AUG-2021, *URLConnection*, *HttpUrlConnection*, *HttpsUrlConnection*, and *Socket*, each in OpenJDK 18 from AUG-2021, *Android WebView 90* from APR-2021, *Google Chrome 92* from AUG-2021, *Microsoft Edge 92* from AUG-2021, and finally, *Mozilla Firefox 91* from AUG-2021. The only HTTP clients that contain closed-source components are Google Chrome and Microsoft Edge, which both rely on the open-sourced Chromium rendering engine, however they have proprietary customizations, *e.g.*, vendor account synchronization.

We can see that support for header fields barely exists among HTTP client libraries, however, the opposite is true for all web browsers: they support all features with only a few exceptions, *e.g.*, for obsolete header fields. The support for versioning information features (`Server`, `X-AspNet-Version`, `X-Powered-By`, `X-Powered-By-Plesk`) remains incomplete across the different softwares: although the header field is accessible in almost all tested API clients, no further routines are available to dynamically react on them, *e.g.*, by disrupting connections to insecure servers. *HttpComponents* has only stubs with no provided logic behind some security-related header fields, however it is one of the few clients that contains code to treat multiple different scenarios of the `Upgrade` field, *e.g.*, a connection protocol upgrade from HTTP to HTTPS, from HTTP/1.1 to HTTP/2, or from HTTP to WebSocket. Furthermore, within the past six months we could see that the number of security-relevant header field stubs has continuously increased in *HttpComponents* and *OkHttp*, *e.g.*, the `Access-Control-Allow-Origin` and the `Strict-Transport-Security` field are now explicitly parsed, although still not evaluated, which has not been the case before. Since *LoopJ* is based on *HttpComponents* it has access to the same feature set, but unfortunately it currently uses an older release with fewer features. We can further see that the HSTS policy has been considered for the *OkHttp* client: there exists a feature request ticket from February 2017 in Square's GitHub product page asking for HSTS support.[7] Unfortunately, although the feature has apparently been implemented in some internal developer builds, it has not yet found its way into the production releases. Finally, the `Socket` class is different from the other contenders, because it is mainly built for low-level network communication that does not consider any information from the ISO/OSI layers four or higher that are required for HTTP headers.

> **Observation**
> The support for security-related header fields is very limited in HTTP client libraries compared to web browsers.

## V. DISCUSSION

We discuss opportunities for the adoption of existing security-related header fields in HTTP client libraries, while we have a look at how a major web browser, *i.e.*, Mozilla Firefox has dealt with these problems. These changes would not only increase the client-side security (*e.g.*, validation of received data), but also the server side security (*e.g.*, less exposure of server configurations).

### A. Recommendations

Of the sixteen security-related header fields that we identified, six are independent of the transmitted payload and every HTTP client can consider the following recommendations.

- **Server/X-AspNet-Version/X-Powered-By/X-Powered-By-Plesk**
  *Purpose:* product advertisement. *Current State:* these header fields are very prevalent in web communication, although such information should not be provided. *Mozilla Firefox:* does not use such information and reports that it should not provide overly detailed information.[8] *Proposed Change:* must preferably not be transmitted at all. HTTP clients should throw exceptions when they encounter such header fields to raise awareness of this issue among the developers. *Security Gain:* a potential attacker has no detailed information about the environment.

- **Strict-Transport-Security**
  *Purpose:* ensures that clients access a certain URL only through secure HTTPS communication channels. *Current State:* numerous HTTP servers support this feature, but HTTP client libraries cannot leverage the provided information. An HTTP client that supports HSTS usually maintains its own HSTS database. *Mozilla Firefox:* fully supports this feature including preload lists. *Proposed Change:* every HTTP client must support this feature and in addition maintain a *centralized* HSTS white-list. *Security Gain:* enhanced protection against man-in-the-middle attacks.

- **Upgrade**
  *Purpose:* can be used to upgrade an established connection to a superior protocol, *e.g.*, from HTTP/1.1 to HTTP/2, or from HTTP to HTTPS. *Current State:* Numerous HTTP servers support this feature, but it is lacking in HTTP client libraries. *Mozilla Firefox:* fully supports this feature. *Proposed Change:* every HTTP/1.1 client must support this feature and in addition should provide support for HTTP/2. *Security Gain:* improved protection against man-in-the-middle attacks.

### B. Proposals

Since the existing headers primarily target web browsers and benefit them, we expect that additional headers specifically designed for HTTP client libraries would be beneficial for them as well. For example, we envisage the development of two new header fields to protect against code execution and data leaks. We provide for each of them a listing that enumerates possible configurations.

- **X-Allowed-Interpretation**
  *Purpose:* declares whether it is safe to process the received data with interpreters for a specific computer language. *Inspiration:* it is similar to the `X-Content-Type-Options` field that prevents MIME type sniffing which could cause arbitrary code execution. *Supported Values:* `any`, `none`, or the name of a computer language, *e.g.*, `HTML` and `SQL`, or multiple comma-separated names. *If no value is provided,* no interpreter must be accepted. *If not used*, the illegitimacy of interpreters is not considered. *Enforcement:* the client must track variable values, *e.g.*, using a dynamic taint analysis framework and ensure that the used interpreters are white-listed. Otherwise the client must report an error and abort the operation. *Benefit:* it reduces the risks associated with arbitrary input data from a web API, *e.g.*, non-sanitized inputs that enable SQL injection attacks.
  *Examples:* We present four typical configurations in Listing 2. The configuration in line two does not add any security and allows every interpretation of the attached data, *e.g.*, it can be used to create SQL queries or to label native UI elements in Java Swing. The configuration in line five prevents code injection attacks and prohibits any interpretation of the attached data, *e.g.*, it cannot be used to create SQL queries or to label native UI elements. The configuration in line eight allows some interpretation and suits data that has to be displayed in a web browser component, *e.g.*, it can be used to create a website, but not to create a SQL query, which effectively mitigates SQL injection attacks. Finally, the configuration in line eleven enables even more interpretation of the attached data and allows the simultaneous use of it in the native UI and web components, *e.g.*, it can be used to label elements in the UI and to create websites, but not to create a SQL query. Similarly, this configuration reduces

---
[7] https://github.com/square/okhttp/issues/3170
[8] https://developer.mozilla.org/en-US/docs/Web/HTTP/Headers/Server

TABLE III
SUPPORT OF HTTP SECURITY-RELATED HEADER FIELDS FOR FRAMEWORKS, JAVA CLASSES, AND WEB BROWSERS. WE USE THE SYMBOL ✔ FOR A FEATURE THAT IS FULLY SUPPORTED; (✔) FOR A FEATURE WITH LIMITED SUPPORT; AND ✘ FOR A NON-EXISTENT FEATURE.

| Header | Glide | HttpComponents | Ion | LoopJ | OkHttp | RetroFit | Volley | HttpsURLConnection | HttpURLConnection | Socket | URLConnection | Android WebView | Google Chrome | Microsoft Edge | Mozilla Firefox |
|---|---|---|---|---|---|---|---|---|---|---|---|---|---|---|---|
| Server | (✔) | (✔) | (✔) | (✔) | (✔) | (✔) | (✔) | (✔) | (✔) | ✘ | (✔) | ✔ | ✔ | ✔ | ✔ |
| X-Powered-By | (✔) | (✔) | (✔) | (✔) | (✔) | (✔) | (✔) | (✔) | (✔) | ✘ | (✔) | (✔) | (✔) | (✔) | (✔) |
| X-Content-Type-Options | ✘ | ✘ | ✘ | ✘ | ✘ | ✘ | ✘ | ✘ | ✘ | ✘ | ✘ | ✔ | ✔ | ✔ | ✔ |
| X-XSS-Protection | ✘ | ✘ | ✘ | ✘ | ✘ | ✘ | ✘ | ✘ | ✘ | ✘ | ✘ | ✘ | ✘ | ✘ | ✘ |
| X-Frame-Options | ✘ | ✘ | ✘ | ✘ | ✘ | ✘ | ✘ | ✘ | ✘ | ✘ | ✘ | ✔ | ✔ | ✔ | ✔ |
| Access-Control-Allow-Origin | ✘ | (✔) | ✘ | ✘ | (✔) | ✘ | ✘ | ✘ | ✘ | ✘ | ✘ | ✔ | ✔ | ✔ | ✔ |
| Strict-Transport-Security | ✘ | (✔) | ✘ | ✘ | (✔) | ✘ | ✘ | ✘ | ✘ | ✘ | ✘ | ✔ | ✔ | ✔ | ✔ |
| Upgrade | ✘ | ✔ | ✘ | ✔ | ✔ | ✘ | ✘ | ✘ | ✘ | ✘ | ✘ | ✔ | ✔ | ✔ | ✔ |
| X-AspNet-Version | (✔) | (✔) | (✔) | (✔) | (✔) | (✔) | (✔) | (✔) | (✔) | ✘ | (✔) | (✔) | (✔) | (✔) | (✔) |
| Content-Security-Policy | ✘ | ✘ | ✘ | ✘ | ✘ | ✘ | ✘ | ✘ | ✘ | ✘ | ✘ | ✔ | ✔ | ✔ | ✔ |
| Access-Control-Expose-Headers | ✘ | (✔) | ✘ | (✔) | ✘ | ✘ | ✘ | ✘ | ✘ | ✘ | ✘ | ✔ | ✔ | ✔ | ✔ |
| Expect-CT | ✘ | ✘ | ✘ | ✘ | ✘ | ✘ | ✘ | ✘ | ✘ | ✘ | ✘ | ✘ | ✔ | ✔ | (✔) |
| X-Powered-By-Plesk | (✔) | (✔) | (✔) | (✔) | (✔) | (✔) | (✔) | (✔) | (✔) | ✘ | (✔) | (✔) | (✔) | (✔) | (✔) |
| Access-Control-Allow-Credentials | ✘ | (✔) | ✘ | (✔) | ✘ | ✘ | ✘ | ✘ | ✘ | ✘ | ✘ | ✔ | ✔ | ✔ | ✔ |
| Timing-Allow-Origin | ✘ | ✘ | ✘ | ✘ | ✘ | ✘ | ✘ | ✘ | ✘ | ✘ | ✘ | ✔ | ✔ | ✔ | ✔ |
| Referrer-Policy | ✘ | ✘ | ✘ | ✘ | ✘ | ✘ | ✘ | ✘ | ✘ | ✘ | ✘ | ✔ | ✔ | ✔ | ✔ |

the attack surface by preventing the interpretation of data in arbitrary languages.

```
// allows every interpretation
X-Allowed-Interpretation: any

// prohibits any interpretation
X-Allowed-Interpretation: none

// only allows the interpretation in JS
X-Allowed-Interpretation: JavaScript

// only allows the interpretation in Swing or JS
X-Allowed-Interpretation: JavaSwing,JavaScript
```

Listing 2. `X-Allowed-Interpretation` configurations

- **X-Allowed-Persistence**
  *Purpose:* it is to specify whether the received data is allowed to be stored persistently on disk. *Inspiration:* the `Content-Security-Policy` field that specifies the trusted sources for specific content. However, this field only specifies from where content can be received, but not where it can be stored locally. *Supported Values:* `any` if all received data can be stored, `none` if data must not be stored, or `only-hashed` if data must be hashed before it can be stored. *If no value is provided,* `none` is used instead. *If not used,* no restriction applies to the data. *Enforcement:* the client must track variable values, *e.g.*, using a dynamic taint analysis framework and ensure that the used storage locations are white-listed. Otherwise the client must report an error and abort the operation. *Benefit:* programs that store data received from a web API will not accidentally leak sensitive information, *e.g.*, by logging a password or saving credentials in a publicly accessible storage location.
  *Examples:* We present three typical configurations in Listing 3. The configuration in line two does not add any security and allows anywhere the storage of the attached data, *e.g.*, it can be stored to a disk or leaked to a console. The configuration in line five provides basic data protection by only allowing hashed data to be persistently stored, which mitigates plain text password leaks. Finally, the configuration in line eight can be used to protect arbitrary sensitive data by preventing the storage of any of the attached data, *e.g.*, it prevents leaking a user name to the console or the disk.

```
// allows to persist the attached data anywhere
X-Allowed-Persistence: any

// only allows to persist hashed data
X-Allowed-Persistence: only-hashed

// prohibits to persist the attached data
X-Allowed-Persistence: none
```

Listing 3. `X-Allowed-Persistence` configurations

## VI. THREATS TO VALIDITY

*Selection of URLs.* The foremost threat to validity of this study is the selection bias, *i.e.*, whether the selection of URLs is representative. The authors of the URL extraction study strived to collect URLs from more than three thousand random apps from various categories in order to obtain relevant results. The analyzed apps were popular and achieved many downloads. Our decision to include URLs that point to the same server, but use a different path or query parameter ensures that every server is accessed, however it might introduce false positives when the same server is serving multiple endpoints.

*Dataset accuracy.* We do not genuinely know the accuracy of the entire URL list that has been constructed using a static analyzer in previous work [11]. The pre-processed list may suffer from inconsistencies due to their rather opportunistic approach, *e.g.*, broken variable assignments, *etc*. However, the URLs generally do not need to be very accurate to retrieve reasonable results; when the domain and the path are correct, the different query parameters generally point to the same server.

*Selection of header fields.* We only considered security header fields that were prevalent in the received HTTP responses and did not scrutinize other header fields. As a result, we missed, for example, the outdated header field `Public-Key-Pins`, which has been used for certificate pinning before `Expect-CT` became available. We only queried the servers with the most commonly used HTTP GET request method, but not other methods such as PUT, DELETE, or POST, which are less prevalent. Moreover, we used the default settings of the *curl* application, where not otherwise specified, to collect the headers, which can alter the results if a server reacts differently depending on the transmitted user agent.

*Library support for security-related header fields.* Where available, we searched in the source code for the support of particular header fields, *i.e.*, we checked whether a header field name exists. Although a match likely indicates that the header field is at least partially supported, its absence does not necessarily indicate that it is unsupported, although it is very likely. We tried to mitigate this risk by intensively investigating the found matches, and by manually skimming relevant classes for such logic.

*Sensitive data.* We assume that every communication involves the transmission of sensitive data and should be secure, which is not always true. However, the detection of data sensitivity is a non-trivial task and not part of this work.

*Lack of validation.* Finally, we did not yet implement the proposed headers and therefore the expected security gain remains speculation. Moreover, the list of proposed headers is not conclusive.

## VII. RELATED WORK

Lavrenovs *et al.* assessed HTTP security header fields in the top one million websites and found that less popular websites tend not to implement security-related features for their users [12]. Notably, they found that nearly 38% of the top one thousand sites implement HSTS, while it is the case for only 17.5% of the top one million HTTPS websites. Buchanan *et al.* performed the same experiment and generally draw very similar conclusions, except that, according to them, *Let's Encrypt SSL* certificates are more than eight times more prevalent in minor sites compared to major sites, because the top sites apparently have the money to buy their own certificates and do not rely on free services [13].

Fahl *et al.* investigated the use of insecure connections in Android apps [14] and found that almost every tenth app is potentially vulnerable to a man-in-the-middle attack. They could capture credentials from various credit cards, social media accounts, web blogs, *etc.*

Adopting new header fields seems to be an appropriate approach to address HTTP issues without breaking existing implementations. Two well-known instances are the cross-site scripting protection proposed by Stamm *et al.* [15] and the protection against downgrading HTTPS connections, which arose from the work of Marlinspike [16]. Both new header fields, *i.e.*, `Content-Security-Policy` and `Strict-Transport-Security`, have become standardized, and they are now supported in major web browsers.

In summary, the work presented in this paper focuses on HTTP header fields adopted in mobile app communications, and it aims to bring attention to the lack of support for header fields in HTTP client libraries.

## VIII. CONCLUSION

We collected the HTTP response header information from 9 714 distinct URLs found in 3 376 Android apps. We discovered that, on average, 93% of the security-related headers are not used in server responses, indicating great potential for future improvements. We also found that unlike major web browsers, the support for such fields in HTTP client libraries is very limited, and that server responses for mobile apps frequently lack them. We encourage a more comprehensive use of existing HTTP headers and timely development of relevant web browser security features in HTTP client libraries. We are developing a technique to enforce the persistence and interpretation policies for input data in mobile apps.

### ACKNOWLEDGMENT

We gratefully acknowledge the financial support of the Swiss National Science Foundation for the project "Agile Software Assistance" (SNSF project No. 200020-181973, Feb. 1, 2019 - April 30, 2022).